\documentclass{phbauth}        
\usepackage{graphicx}

\begin{document}

\begin{frontmatter}

\title{Relative evaporation probabilities of $^3$He and $^4$He from the surface of superfluid $^4$He}

\author[address1]{Jonathan\,P.\,Warren} and
\author[address1]{Charles\,D.H.\,Williams\thanksref{thank1}},

\address[address1]{School of Physics, University of Exeter, Stocker Road,
Exeter EX4 4QL, United Kingdom}

\thanks[thank1]{Corresponding author. E-mail: c.d.h.williams@ex.ac.uk} 

\begin{abstract}
We report a preliminary experiment which demonstrates that $^3$He atoms in Andreev states are evaporated by high-energy ($E/k_\mathrm{B}\approx10.2\,\mathrm{K}$) phonons in a quantum evaporation process similar to that which occurs in pure $^4$He. Under conditions of low  $^3$He coverage, high-energy phonons appear to evaporate $^3$He and $^4$He atoms with equal probability. However, we have not managed to detect {\it any} $^3$He atoms that have been evaporated by rotons, and conclude that the probability of a roton evaporating a $^3$He atom is less than 2\% of the probability that it evaporates a $^4$He atom.
\end{abstract}


\end{frontmatter}



When small quantities of $^3$He are added to bulk superfluid $^4$He below $T{\sim}100\,\mathrm{mK}$ the atoms occupy so-called Andreev states \cite{ANDREEV} and form a degenerate two-dimensional fermion system. It has previously been reported that $^3$He can be evaporated by phonons \cite{BAIRD} in a quantum evaporation \cite{CDHW} process. This report briefly describes an experiment to compare the evaporation probabilities for $^3$He and $^4$He atoms by positive group-velocity rotons and high-energy phonons.
\par
\begin{figure}[b]
\begin{center}\leavevmode
\includegraphics[width=1.0\linewidth]{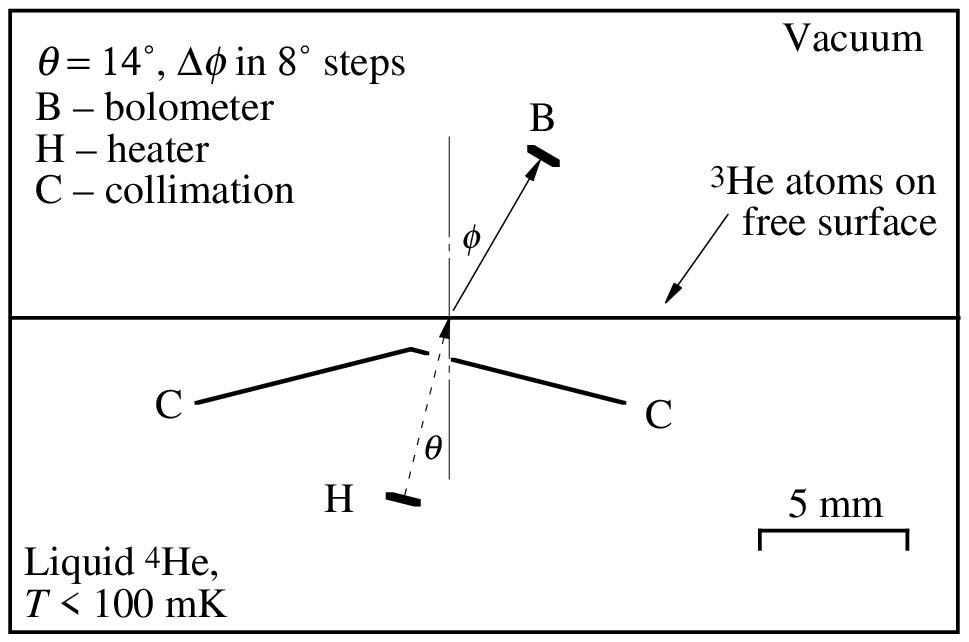}
\caption{Schematic diagram of the experiment. The bolometer angle $\phi$ is adjusted with a stepper-motor.}\label{HARDWARE}\end{center}\end{figure}

An electrically heated $1\,\mathrm{mm}^2$ thin-film heater in bulk superfluid $^4$He was used to generate excitations which travelled ballistically (path length $6.3\,\mathrm{mm}$ at $\theta=14^\circ$ to the vertical) to the free surface (Fig.\,\ref{HARDWARE}). Evaporated atoms were detected with a constant-temperature superconducting bolometer\cite{BOLO} at angle $\phi$ to the vertical and at radius $5.8\,\mathrm{mm}$ from the point of intersection of the liquid surface with the centre of the excitation beam.

\par
The high-energy phonons that participate in evaporation have a narrow energy-distribution which peaks at $E/k_\mathrm{B}=10.2\,\mathrm{K}$ and cuts off below $10\,\mathrm{K}$ \cite{PHONONS-EX}. These phonons are generated about a millimeter in front of the heater by a complicated up-scattering mechanism \cite{PHONONS-TH}. The $^3$He atoms are less tightly-bound to the liquid surface than the $^4$He and therefore have a shorter time of flight to the bolometer \cite{BAIRD}. The signals were recorded as a function of angle and surface concentration of $^3$He. The signal shapes are complicated because the $^3$He affects the bolometer responsivity and time-constant, but a simplified analysis is possible by considering only energy-conserving processes involving the dominant $10.2\,\mathrm{K}$ phonons, and the relative probabilities of phonon--atom evaporation processes can be inferred as follows.

\begin{figure}[b]
\begin{center}\leavevmode
\includegraphics[width=1.0\linewidth]{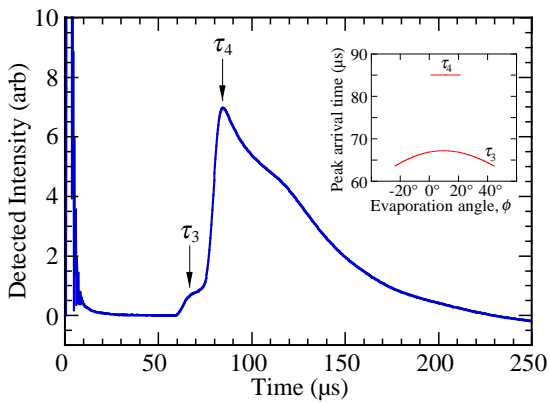}
\caption{A phonon--atom evaporation signal taken at $\phi=7^\circ$ and $n_\mathrm{3S}=1.1\,\mathrm{nm}^{-2}$. The inset shows predicted arrival times associated with evaporation by a 10.2\,K phonon.}\label{SIGNAL}\end{center}\end{figure}

\par
\begin{figure}[t]
\begin{center}\leavevmode
\includegraphics[width=1.0\linewidth]{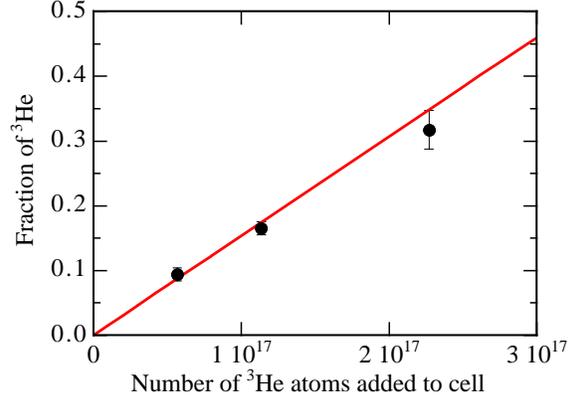}
\caption{Measured fraction of total evaporation signal due to $^3$He atoms (points) compared with calculated coverage of surface by $^3$He (line).}\label{COVERAGE}\end{center}\end{figure}

With an isotopically pure $^4$He surface, the measured time $\tau_4$ of the peak
in the phonon-$^4$He evaporation signal at $\phi=11^\circ$ (the kinematic angle
of evaporation for $10.2\,\mathrm{K}$ phonons) is used to establish the arrival
time of these phonons at the surface (Fig.\,\ref{SIGNAL}). Next, the arrival time
$\tau_3(\phi)$ of a $^3$He atom evaporated by a $10.2\,\mathrm{K}$ phonon as a
function of evaporation angle $\phi$ is calculated (inset to Fig.\,\ref{SIGNAL}).
The $^3$He atoms have a two-dimensional Fermi distribution of momentum. Those
nearest the Fermi energy have the earliest arrival times and trvel along paths at
the extremes of the angular-distribution of allowed evaporation directions. At a
given bolometer position $\phi$, the amplitude of the bolometer signal
$S(\phi,t)$ at time $t=\tau_3(\phi)$ is -- ignoring minor errors due to imperfect
collimation -- entirely due to $^3$He atoms which have been evaporated by
$10.2\,\mathrm{K}$ phonons. By integrating the measurements of
$S(\phi,\tau_3(\phi))$ and $S(\phi,\tau_4)$ over the solid angle, the relative
numbers of $^3$He and $^4$He atoms evaporated by $10.2\,\mathrm{K}$ phonons can
be estimated. We conclude that when $^3$He coverage is below 0.5 monolayers
($3.3\,\mathrm{nm}^{-2}$) a phonon will evaporate an atom of either isotope with
equal probability (Fig.\,\ref{COVERAGE}).

\par

In the light of this result we were surprised to find no evidence, despite a
careful search, that positive group-velocity rotons can quantum evaporate
$^3$He atoms. We believe that the probability that a roton evaporates a $^3$He
atom is less than our detector noise limit, {\it i.e.} 2\% of its probability of
evaporating a $^4$He atom.



\end{document}